# Human-Centered Design and Evaluation of a Workplace for the Remote Assistance of Highly Automated Vehicles

Andreas Schrank [a*], Fabian Walocha [a], Stefan Brandenburg [b], and Michael Oehl [a]

[a]*Institute of Transportation Systems, German Aerospace Center (DLR), Braunschweig, Germany;* [b]*Cognitive Psychology and Human Factors, Chemnitz University of Technology, Chemnitz, Germany*

* Corresponding author:

Andreas Schrank

German Aerospace Center (DLR)

Institute of Transportation Systems

Lilienthalplatz 7

38108 Braunschweig

Germany

Telephone +49 531 295 1015

andreas.schrank@dlr.de

Andreas Schrank is a Human Factors researcher at the German Aerospace Center (DLR), Institute of Transportation Systems. He obtained his Master's degree in Psychology from Heidelberg University, Germany, while also studying at Şehir University in Istanbul, Turkey, and at the University of North Carolina at Chapel Hill, NC, USA. In his research, Andreas Schrank focuses on human-machine interaction in the realm of automated driving with an emphasis on the Human Factors of remotely operated and automated vehicles, particularly regarding roles, tasks, systems, and user-centered interfaces. He is involved in the major EU research project "Hi-Drive" as well as several German national research projects on this topic.

Fabian Walocha is a researcher at the German Aerospace Center (DLR), Institute of Transportation Systems. He obtained his Master's degree in Computer Science from Université Jean Monnet in Saint-Étienne, France. His research interests revolve around human-machine interaction, user-adaptive systems and user state modeling.

Prof. Stefan Brandenburg (PhD in 2014) studied Psychology and Philosophy at Technische Universität Chemnitz, Germany, and The University of Oklahoma, OK, USA. He obtained his PhD from Technische Universität Berlin in Cognitive Ergonomics in 2014. Since 2023, he is Professor of Cognitive Psychology and Human Factors at Technische Universität Chemnitz, Germany. His research topics include the design of driver-vehicle interactions for automated vehicles, usability and user experience design, and ethical aspects of (new) technology.

Dr. Michael Oehl is head of the research group Human-Machine Interaction (HMI) at the German Aerospace Center (DLR), Institute of Transportation Systems in Braunschweig and Berlin, Germany. The research group is devoted to user-centered human-machine interaction and interface design in future transportation systems. He is an expert in the ISO and German DIN with regard to remote control of automated vehicles. Additionally, he is adjunct senior lecturer for Traffic Psychology at the German Police University. Michael Oehl has a strong background in Human Factors as well as Engineering and Traffic Psychology. He studied psychology at the University of Konstanz, RWTH Aachen University and received his PhD from Leuphana University of Lüneburg. Among other scholarships, he was a Japan Society for the Promotion of Science (JSPS) fellow at the Cognitive Systems Engineering Lab at the University of Tokyo.

# Human-Centered Design and Evaluation of a Workplace for the Remote Assistance of Highly Automated Vehicles

Remotely operating vehicles utilizes the benefits of vehicle automation when fully automated driving is not yet possible. A human operator ensures safety and availability from afar and supports the vehicle automation when its capabilities are exceeded. The remote operator thus fulfills the legal requirements in Germany as a Technical Supervisor to operate highly automated vehicles at SAE 4. To integrate the remote operator into the automated driving system, a novel user-centered human-machine interface (HMI) for remote assistance workplaces was developed and initially evaluated. The insights gained in this process were incorporated into the design of a workplace prototype for remote assistance. This prototype was now tested in the study reported here by 34 participants meeting the professional background criteria for the role of Technical Supervisor according to the German law by using typical remote assistance scenarios created in a simulation environment. Even under elevated cognitive load induced by simultaneously engaging in a secondary task, participants were able to obtain sufficient situation awareness and quickly resolve the scenarios. The HMI also yielded favorable usability and acceptance ratings. The results of the study inform the iterative workplace development and further research on the remote assistance of highly automated vehicles.

## 1    Introduction

The automation of driving technology advances quickly. It is associated with benefits concerning safety, reliability, and passenger comfort, as well as the reduction of economic and environmental costs of mobility (Litman, 2020; Schoitsch, 2016). It is considered key for the fundamental shift in transportation away from individual mass motorization to flexible on-demand mobility solutions, for instance, shuttle vehicles (Iclodean et al., 2020). However, highly automated driving (SAE level of automation 4; Society of Automotive Engineers, 2021) in urban mixed-traffic environments will remain challenging for automated vehicles. On this level of automation, situations may occur that the vehicle's automated driving system cannot handle (Kalisvaart, 2021). Human operators are able to tackle even unforeseen situations with creativity and ingenuity, and can be fruitfully included into automated transportation systems consisting of a highly automated vehicle (HAV) and a remote operator (RO). In remote operation systems, an RO oversees vehicle operations from a control center. The RO overviews and analyzes traffic situations that automated vehicles encounter. They provide guidance to the vehicle automation on how to tackle difficult situations. This is conceivable for any vehicles with high driving automation (SAE 4 or higher), shuttle buses, and other vehicles, e.g., personal vehicles, transportation vehicles such as vans and trucks, as well as larger buses. Remote operation could, therefore, help to overcome situations that the automation alone cannot handle, resulting in safer and smoother operations of HAVs. A pivotal component of a safe and smooth HAV remote operation system is the RO's workplace. The following paper will describe the design for a conceptual prototype for a workplace for remote assistance, a variant of remote operation, and its user evaluation focusing on the central indicators performance, situation awareness, and workload.

## 1.1 *Workplaces for Remote Operators*

ROs will be a core component of HAV remote operation systems. The human-machine interface (HMI) of the RO workplace is essential for safe, effective, and efficient operations. Remote operation can mainly be implemented in two different ways. First, in the *remote driving* approach, also known as direct or teleoperated driving, the RO executes the dynamic driving task (DDT) including braking, steering, and accelerating in real time (Society of Automotive Engineers, 2021). The input given resembles manual driving, requiring the RO's continuous attention. Second, the *remote assistance,* or indirect, approach is defined as "event-driven provision, by a remotely located human, of information or advice to [… a] vehicle in driverless operation in order to facilitate trip continuation when the ADS [automated driving system] encounters a situation it cannot manage" (Society of Automotive Engineers, 2021, p. 18). The HMI presented and evaluated in this paper aims to enable remote assistance at Level 4 Automation (Society of Automotive Engineers, 2021). Hence, the HMI is not designed for executing the DDT. In accordance with the definition of Level 4, the remote assistant who oversees an HAV does not serve as a fallback for the automation. The HAV must be able to transfer itself into a minimal-risk state, posing the least danger possible to itself, its passengers, as well as surrounding road users. Remote assistance in this implementation is the only permissible way of implementing remote operation of vehicles on public roads in Germany to this date (StVG § 1e, 2021/12.07.2021).

During remote assistance, the RO's main task is the processing of requests for assistance coming from the supervised HAV (see Figure 3). According to German law, ROs specified as Technical Supervisors ("Technische Aufsicht") are responsible to check and assist an HAV based on evidence that it requires support ("Evidenzkontrolle"). This means that an RO becomes involved only when the vehicle detects an event that it cannot handle autonomously and thus submits a request for assistance to the RO (StVG § 1e, 2021/12.07.2021).

Even though German law demands that interventions by the RO cannot be time-critical, (a) *task reaction time*, i.e., the duration passed from the request's appearance on the RO's workplace HMI to the RO's acceptance of the request, is still considered a key performance indicator as it is essential for efficient operations and therefore relevant for the economically feasible implementation of RO systems. In addition, (b) *task completion time*, i.e., the time passed from the RO's acceptance of the request to the resolution of the task, is an indicator to measure how long it took an RO to resolve a task.

The literature on workplace HMIs for remote operation is scarce. Following a human-centered design process, Kettwich et al. (2021) designed and evaluated a click-prototype for a remote operation workplace HMI. It was tailored to the remote assistance of SAE 4 shuttle buses from a public transport control center. Apart from this research, although software and hardware solutions for the remote operation of vehicles already exist (e.g., DriveU.auto, 2023; Herger, 2023; T-Systems, 2023; Vay, 2022), no systematic research has been conducted in a highly controlled laboratory environment to develop and evaluate a prototypical HMI for HAV *remote assistance* to the authors' knowledge. Remote assistance here is defined as the task of the Technical Supervisor according to the current German Law for Autonomous Driving. In particular, there is a gap in research on workplace HMIs for remote operation of vehicles in the contexts of public transport, logistics, and individual mobility that are tailored to the needs, expectations, and operation styles of control centers in these areas. Therefore, the goal of this work is the user-centered design of a prototypical workplace HMI for a concrete implementation of remote operation, remote assistance, and its evaluation regarding performance, situation awareness and workload in routine remote assistance tasks. Also, we want to assess the operator's subjective experience by assessing their ratings of usability, user experience, and acceptance.

## *1.2 Situation Awareness*

Similar to a driver, a remote operator (RO) needs to perceive and identify relevant elements of a traffic situation. They must integrate them to a coherent understanding of the situation and be able to predict how relevant elements will change in the future.

These operator tasks can be described by situation awareness (SA). The hierarchical SA model of Endsley ((1995) proposes three levels of SA. A lower level of SA needs to be fulfilled in order to reach a higher one. On SA level 1, a RO has to *perceive* characteristics of the traffic environment like road layout and condition, traffic signs, and other road users. On SA level 2, the RO has to analyze and integrate these elements in accordance with their goals to "form a holistic picture of the environment, *comprehending* the significance of objects and events" (Endsley, 1995, p. 37). For example, a pedestrian crossing an HAV's lane is relevant to the RO's goal to continuously drive on this lane. On SA level 3, the RO *predicts* how the situation will unfold. A result of high SA is that the RO commands the HAV change lanes in order to avoid the predicted collision.

In a remote setting, it may be difficult to achieve high levels of SA because ROs cannot perceive the elements of the driving situation directly and without delay (SA level 1), or react immediately to them (based on SA levels 2 and 3). Also, there is no direct link between a RO and the surrounding traffic environment. Information of the driving situation is sensed via technology, transmitted to the RO's workplace, and displayed to them through the interface. Similarly, the RO's reaction is mediated through data transmission, in-vehicle processes, and execution by actuators, causing delays between operator inputs and vehicle reactions as well as vehicle actions and presented status. Decoupling action, perception, decision, and reaction by inserting intermediate steps of deconstruction, transmission, and reconstruction into the process has important implications: Distortions may occur in any of these steps, negatively impacting the RO's SA (Tittle et al., 2002). For instance, Darken et al. (2001) stated that participants performed poorly in spatial orientation as well as object identification tasks when video feedback was supplied to remote observers. Thus, the HMI design of the RO workplace concerning the selection of information modes (visual, auditory, etc.) and the way information is displayed to the RO affects their level of SA (Endsley, 1995; Endsley et al., 2003; Hollands et al., 2019). As a result, the RO's workplace needs to ensure high levels of SA.

## *1.3 Workload*

Workload is the experienced difference between required and supplied information processing capability (Hart & Staveland, 1988). It is associated with task performance. Therefore, a workplace for remote operation should balance the requirements of tasks to avoid overload, which leads to stress, or underload, which is associated with boredom (Wickens, 1984).

In workplace design, all tasks, be they primary or secondary, need to be considered in workload assessment. Especially secondary tasks impose cognitive load on operators (Sweller, 1988), thereby increasing perceived workload. These tasks can be (a) directly relevant to fulfil the primary task, for example when additional pieces of information need to be gathered from other sources. They can also be (b) indirectly relevant as part of other responsibilities of an operator, e.g., an incoming request for support by an HAV while already processing another HAV's request. However, they can also be (c) irrelevant to an operator's responsibilities, i.e., distractions. An example of the fatal consequences of being distracted from job-related tasks is the rail disaster of Bad Aibling in Bavaria, Germany. A train controller distracted himself from his rail traffic management task by playing a game on his phone, leading to a collision of two trains on a single-track stretch, killing twelve (British Broadcast Corporation, 2016).

In Human Factors research, examining the impacts of a secondary task on an operator's workload has a long tradition (e.g., Ogden et al., 1979). In the case of a RO's task set, generic cognitive secondary tasks such as the n-back task (Kirchner, 1958) can be used as proxies for cognitive load that might result from tasks that the RO could have, like the RO's parallel assistance of several HAVs. The n-back task is widely used in driving-related studies to systematically vary workload (Pfannmüller et al., 2015; Reimer & Mehler, 2011; Wu et al., 2019).

Hence, workplaces for ROs should be designed so that primary and secondary tasks do not lead to an increase of the operator's workload that would severely deteriorate performance. This is especially important as processing multiple tasks at the same time affects the operators' workload and their SA. In these situations, operators need to keep multiple pieces of information in their working memory, leaving less cognitive resources for gaining high levels of SA (cf. Baumann et al., 2008).

To summarize, an HMI for remote operation needs to be designed to enable effective and efficient operations, to balance the RO's workload, and to ensure their SA. In addition, user-focused variables need to be considered.

## 1.4  *Usability, User Experience, and Acceptance*

The user's subjective usability is crucial for their smooth interaction with technical systems. The perceived usability is relevant because it determines how well the user is able to access information from the system and interact with it. High subjective usability is achieved when the interaction between user and system is effective, efficient, and satisfying (International Organization for Standardization, 2018). User experience is a concept that assesses how satisfied users are when interacting with a system (Hassenzahl, 2008; Minge et al., 2017). It is inevitable for developing successful user-centered products (Schrepp et al., 2017a). Finally, user acceptance is imperative for the success of newly introduced technology as it determines whether a new technology will be adopted by its designated user group (van der Laan et al., 1997). All these concepts are of utmost importance when it comes to workplace design as they directly influence efficient, effective, and safe operations.

## 1.5  Research Objectives and Hypothesis

The goal of this work is the user-centered design of an HMI for remote assistance following the established guidelines and its evaluation regarding performance, situation awareness SA and workload. Also, we want to assess the participants' perceived usability, ratings usability, user experience, and acceptance when interacting with the HMI. To achieve this goal, three research objectives were examined in the study.

The first objective examined whether participants show lower performance at increasing levels of cognitive demand in routine remote assistance tasks using the proposed workplace HMI for remote assistance. The overall hypothesis was:

- H1 (performance): When the level of cognitive demand increases, participants' performance decreases.

It separates into three sub-hypotheses:

- H1.1 (task reaction time): When the level of cognitive demand increases, participants require more time to react to an incoming notification which manifests in more time passed from the appearance to the acceptance of the notification.
- H1.2 (task completion time): When the level of cognitive demand increases, participants require more time to process a task which manifests in more time passed from the acceptance of the notification to the completion of the task.

- H1.3 (number of correct n-back comparisons): When the level of cognitive demand increases, participants' numbers of correct n-back comparisons decreases.

The second objective tested whether participants report lower SA at increasing levels of cognitive demand routine while processing remote assistance tasks using the proposed workplace HMI for remote assistance. The corresponding hypothesis was:

- H2 (subjective SA): When the level of cognitive demand increases, participants' reported SA ratings decrease.

The third objective examined whether participants report higher workload with increasing levels of cognitive demand while processing remote assistance tasks using the proposed workplace HMI for remote assistance. Here, the hypothesis was:

- H3 (subjective workload): The participants' reported ratings of workload increase with increasing levels of cognitive demand.

In addition to these objectives, we examined the participants' ratings of usability, user experience, and acceptance. Thereby we wanted to gain first insights on the participants' subjective experience with the remote assistance workplace. Our analysis examined how participants assess the usability of the presented HMI for remote operation and how they rate their satisfaction.

## 2 Method

### 2.1 Sample

Participants were acquired through postings in buildings and online platforms of engineering departments of universities and research centers in Germany. The participants volunteered but were compensated monetarily for their participation with 25 euros. The study was conceptualized and realized in accordance with the Declaration of Helsinki. The institutional review board of the research institution in which the study was conducted approved of the study. Informed consent was obtained from all participants before the experiment. The participants were allowed to stop the study at any point without justification or consequence.

Of the $N = 41$ participants who took part in the study, seven had to be excluded due to issues in the data collection process. Technical issues in the tools used for the collection of either questionnaire or performance data rendered some data unusable in these participants. Only participants with complete datasets ($N = 34$) were included in the analysis. The final sample analyzed consisted of 34 participants (four female). Participants' ages ranged from 23 to 31 years ($M = 26.2$, $SD = 2.31$). 62% of the participants had experience in monitoring technical systems such as airplanes, automated vehicles, wind channels, agricultural robots, pumps, and machines. Their affinity for technology (Franke et al., 2019) was high ($M = 4.94$, $SD = 0.48$; scale poles 1: low – 6: high). All participants had normal or corrected-to-normal vision and possessed a valid driver's license for passenger vehicles. To comply with legal requirements, only participants with a university or state-certified technician degree in the following disciplines were allowed: mechanical, automotive, electrical, aerospace, and aviation engineering (StVG, 2021). 21 participants (62 %) held a Bachelor's degree as their highest academic degree, thirteen a Master's degree. More than a third (35%) of the participants stated to drive a vehicle multiple times per month, about 29% reported to drive several times a week. All participants had heard about HAVs in the past. 91% expressed interest or strong interest in HAV technology indicated by responding with values 4 or 5 on a Likert scale on interest in AVs (1: "not interested at all" – 5: "very interested", *M* = 4.29, *SD* = 0.72). 28 participants (82%) indicated not to have used HAVs so far.

## *2.2 Experimental Design*

The experimental design was a 3×3 within-subjects design. The independent variables were the primary task (Scenarios 1-3) and the secondary task to induce additional cognitive load (none, 1-back, 2-back). Dependent variables were performance in primary and secondary task, workload, SA, usability, user experience, and acceptance (see sections 2.3.4 and 2.3.5).

## *2.3 Materials*

### *2.3.1 Primary Task (Scenarios)*

Three scenarios were used as primary tasks in the study. Figure 1 displays a screenshot of each scenario. The scenarios were extracted from a previously compiled catalogue of scenarios in remote operation (Kettwich et al., 2022) because they were considered typical for routine tasks in remote assistance. The steps of the interaction between RO and workplace are listed comprehensively in Table A1 in the appendix.

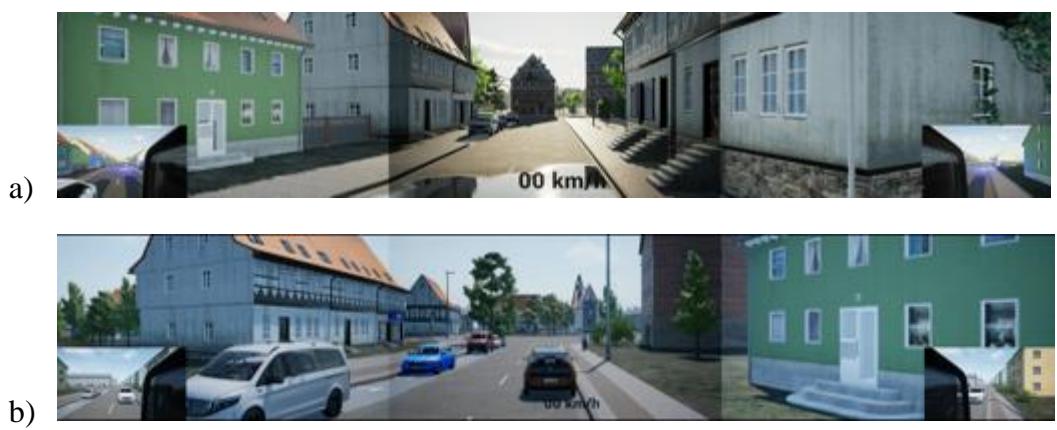

a)

b)

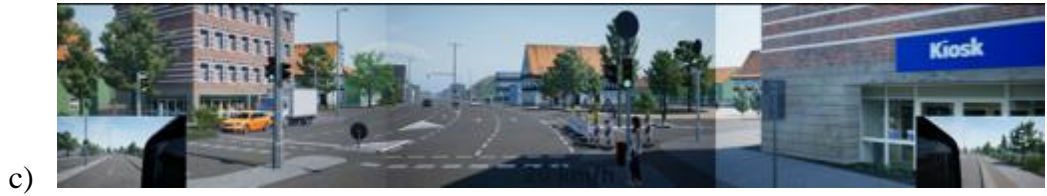
c)

Figure 1. Screenshots of simulated scenarios used in the study. The differences in illumination appear more pronounced here than on the prototypical workplace since separate screens were used to display the images, putting less emphasis on these differences. (a) Scenario 1: Detected Situation Unclear, (b) Scenario 2: Blocked Lane, and (c) Scenario 3: Rerouting.

*Scenario 1: Detected Situation Unclear.* In this scenario, the supervised HAV detects an obstacle on the road, stops and reports the incident to the RO. The detected obstacle is a puddle on the road which reflects the surrounding buildings, so the automation is uncertain whether the vehicle can continue its ride. The RO observes the situation via the supervised HAV's on-board cameras (transmission of video images) and gives clearance so the vehicle can continue its journey. After assessing the situation, the primary task for the RO therefore is giving clearance to continue driving.

*Scenario 2: Blocked Lane.* A vehicle is parking on the lane that the supervised HAV uses, blocking the lane and disabling the HAV from continuing its ride. The HAV stops and provides a corresponding message to the RO. The RO checks the situation on site via the HAV's cameras and sets waypoints for an alternative trajectory using the lane for oncoming traffic to bypass the parking vehicle. The primary task for the RO is to set waypoints to calculate a new trajectory.

*Scenario 3: Rerouting.* Because of a road closure, the supervised HAV needs to change its route. The RO views the road closure via the HAV's cameras and suggested alternative routes and chooses one of them. Thus, the RO's primary task is selecting one of several proposed routes presented on the touchscreen.

*2.3.2     Secondary Task*

As a secondary task, the n-back task (Kirchner, 1958) was included. Its purpose was to modulate the RO's cognitive load in order to simulate phases of elevated workload that are likely to occur in the RO's work. In this task, participants had to compare a presented digit with the digit presented *n* steps before the current one. The higher the *n*, the more digits had to be retained in the participants' working memory, increasing their workload. The n-back task was presented visually and auditorily on a tablet computer distinct from the investigated workplace HMI (Figure 2, bottom right). However, participants were instructed to listen to the auditory presentation only and give their response verbally. The experimenter assured that participants followed these instructions. From a list of 30 digits plus *n,* a single digit from one to nine was played auditorily using the tablet computer's speaker and displayed visually on the screen of the tablet computer every five seconds, so that a total of 30 n-back comparisons had to be made per trial. The order of digits was determined for each trial by randomly assigning one out of four lists that contained a specific order of digits. Participants were instructed to respond verbally with "correct" if the digit presented *n* steps before the current one was identical with the current one, and to respond with "incorrect" if the digit presented *n* steps before the current one differed from the current one. Participants were asked to respond before the next digit was presented, i.e., they had to respond within less than five seconds. The experimenter logged the participants' responses.

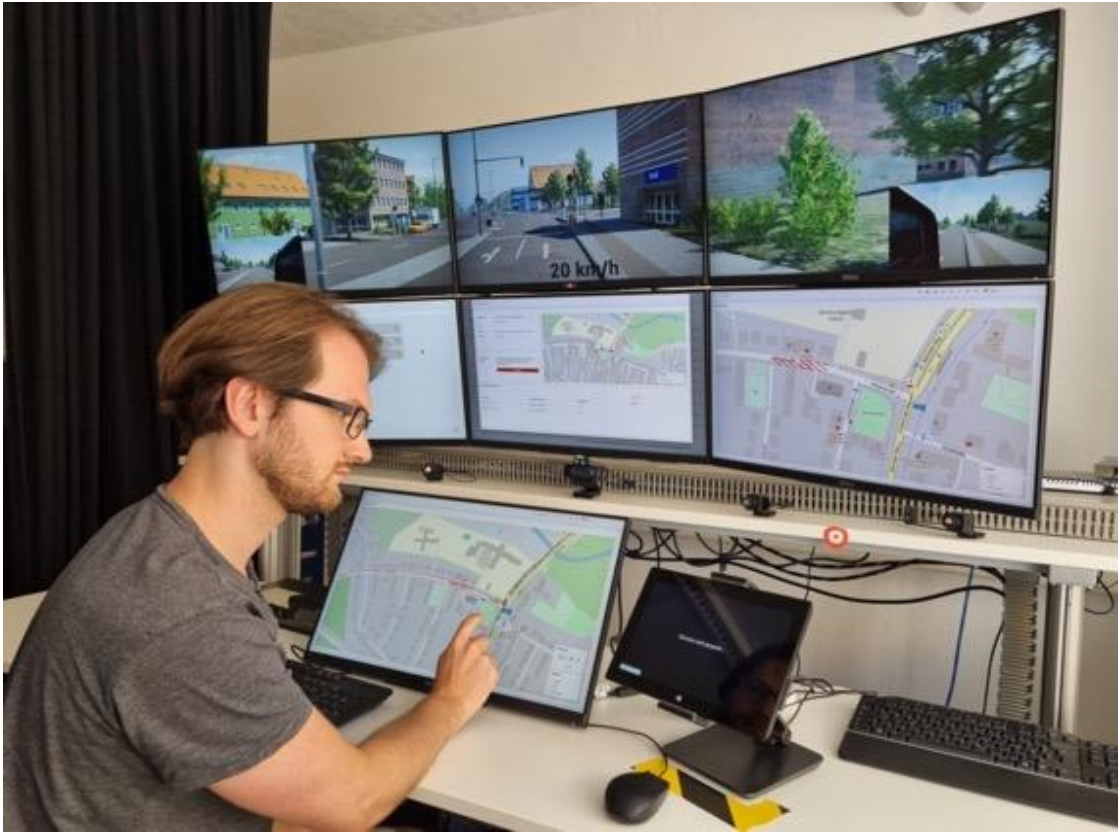

Figure 2. The prototypical workplace for remote assistance and a tablet computer to present the n-back task's stimuli (small dark screen on the right-hand side of the touchscreen). The top row of screens shows the video streams from the on-board cameras of the simulation. The second row of screens presents information on the technical state of the supervised HAV (left screen), an overview of vehicle requests and their respective status (center), and a map of the environment surrounding the supervised vehicle (right). The touchscreen on the bottom is used by the RO to interact with the supervised vehicle, e.g., by setting waypoints or selecting alternative routes on a map, depending on the scenario.

### 2.3.3 Human-Machine Interface (HMI)

The structure and components of the HMI for remote assistance strongly resembled the click-prototype presented and evaluated in Kettwich et al. (2021). It consisted of seven screens of which six were regular computer monitors (24'' Dell, 16:9 ratio), set up in two rows with three monitors each, and another monitor with the same specifications but including a touch feature (see Figure 2). The elements of the HMI as well as the interaction design are described in detail by Kettwich et al. (2021). The steps of the interaction between RO and the supervised HAV are depicted in Figure 3.

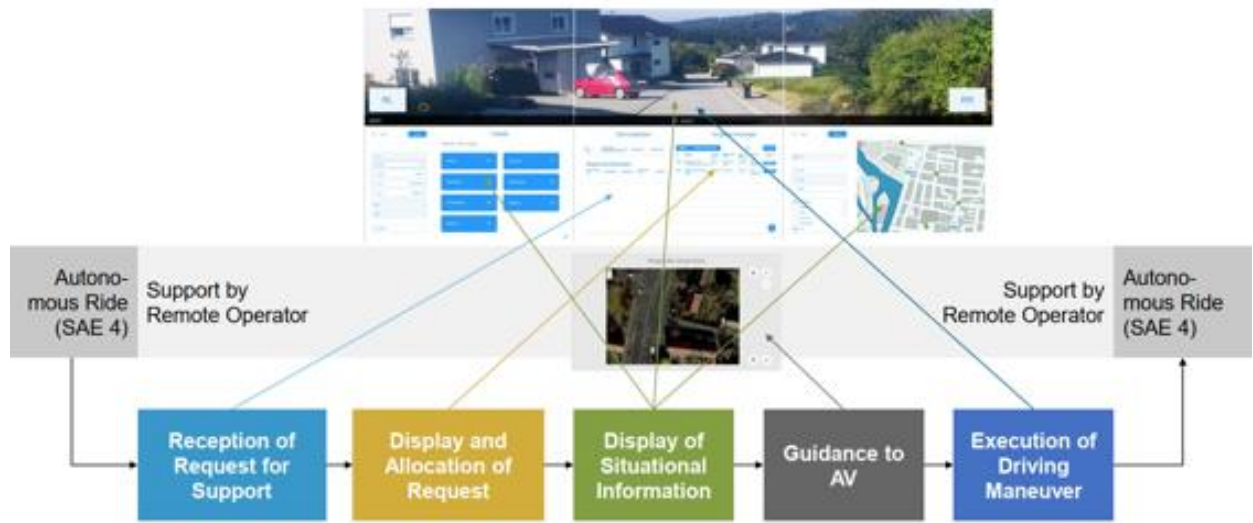

Figure 3. Interaction between remote operator and the supervised HAV between phases of autonomous driving.

In all three scenarios (see Figure 1), the HAV drove in highly automated mode before noticing that it needed the RO's support. Subsequently, it submitted a request to the RO's workplace. The operator received the request for support on the central screen of the second row from the top in the section for incoming notifications that also included some core information such as the HAV's ID, the issue that requires the RO's support, and the spatial position of the HAV. By clicking on "Accept", the RO could allocate the task to themselves, transferring the request to a table containing current tasks. Here, further details such the latest video stream from the HAV, its position on a map and details regarding its technical state, were displayed. Furthermore, a suggestion for an action, such as "Give Clearance", "Set Waypoints", or "Select Alternative

Route", were provided. This information supported the RO's decision on how to assist the HAV. Finally, the RO's input was transmitted to the HAV and executed before it returned to the highly automated driving mode.

*2.3.4     Objective Measures*

We collected three measures that quantified the participants' performance, two of them regarding the primary task, one regarding the secondary task.

*Measures of Primary Task.* Regarding the primary task, the objective was to examine how fast participants were able to react to incoming notifications. Even though remote operation must not be time-critical by law, from an economic point of view, a speedy reaction is still favorable to enable a business case built on remote operation. In addition, the duration participants spent for completing the task was measured to investigate if the HMI is suitable to fulfil the RO's task in a timely manner. Hence, we measured the participants' performance in the primary task using two variables: (a) the time that passed from appearance to the acceptance of the support request in seconds, hereafter called *task reaction time*, and (b) the time that passed from acceptance to completion of the support in seconds, called *task completion time*.

*Measure of Secondary Task.* To measure how much cognitive load was induced, we measured the participants' performance in the secondary task using the number of correct n-back comparisons (max. 30) in the n-back task (see section 2.3.2 for further details).

*2.3.5     Questionnaires*

In addition, we collected self-report data using six questionnaires.

*NASA-TLX.* The NASA Task Load Index (NASA-TLX; Hart & Staveland, 1988) was used to measure subjective workload after each trial. It is an established multi-dimensional measure for participants to report how taxing they experienced a task. The questionnaire distinguishes between six dimensions of workload: mental demand, physical demand, temporal demand, performance, effort, and frustration. Responses to each item were collected on a 7-point Likert scale ranging from the poles 1: "low" to 21: "high".

*SART.* The Situation Awareness Rating Technique (SART; Taylor, 1990) assessed the participants' situation awareness (SA) post-trial. It was originally developed to determine pilots' SA and consists of three subscales: demands on attentional resources, supply of attentional resources, and understanding of the situation. Responses to each item were collected on a 7-point Likert scale. The poles depended on the specific item but ranged from a low to a high degree on a specific construct. For instance, the poles of the item "instability of the situation" were 1: "The scenario is entirely stable" to 7: "The scenario is entirely unstable". An overall SART score was calculated by deducting the difference of attentional demand and attentional supply from understanding.

*SUS.* The Systems Usability Scale (SUS; Brooke, 1996) measures perceived usability. Originating from the need to quickly evaluate usability in software development, it is an economic solution to assess the construct robustly and across a wide range of domains (Bangor et al., 2008). Responses to each item were collected on a 5-point Likert scale ranging from the poles 1: "I do not agree at all" to 5: "I totally agree". A single indicator value between 0 and 100 summarizes the status of the investigated HMI regarding the participants' impression how well it was suited to execute a particular task. As a global assessment tool, SUS was administered at the end of the study.

*UEQ-S.* The User Experience Questionnaire Short Version (UEQ-S; Schrepp et al., 2017b) assesses user experience. It consists of two subscales, the pragmatic and the hedonic subscale. While the pragmatic one captures a construct that leans towards usability, the hedonic one focuses on the emotional quality of the interaction. The UEQ-S is a condensed version of the standard UEQ (Schrepp et al., 2017a), compressing the six subscales from the standard UEQ to the beforementioned two subscales. Using the UEQ-S ensured a more economic collection of subjective data on participants' emotional experience with a system. Responses to each item were collected on a 7-point Likert scale. The poles were semantically opposed statements on a construct each, e.g., 1: "obstructive" to 7: "supportive".

*Acceptance Scale.* The Acceptance Scale (van der Laan et al., 1997) was developed as a standard tool to measure driver acceptance of new technology. With nine items divided in two scales, it measures the usefulness of a system, associated with usability, and the user's satisfaction with said system, similar to user experience. Thus, it is conceptually related to the structure of UEQ-S but adds the dimension of user acceptance when both subscales are considered holistically. Responses to each item were collected on a 5-point Likert scale. The poles were semantically opposed statements on a construct each, e.g., 1: "useful" to 5: "useless".

*ATI.* The Affinity for Technology Interaction Scale (ATI; Franke et al., 2019) assessed the participants' affinity for technology. This construct entails a person's tendency to engage in interaction with technology. With its satisfying psychometric characteristics, the ATI scale measures the affinity for technology with nine items. In this study, the construct was used to describe the sample. Responses to each item were collected on a 6-point Likert scale ranging from the poles 1: "I completely disagree" to 6: "I completely agree".

*2.4 Procedure*

An overview of the procedure is given in Figure 4. First, the experimenter welcomed participants, briefed them about the objectives of the study and asked them to sign an informed consent, a non-disclosure agreement and a data protection declaration. Subsequently, participants filled in the sociodemographic questionnaire and completed the ATI scale (Franke et al., 2019) before they received a detailed explanation of the research context. Participants were instructed to imagine being a remote operator who assists HAVs that operate as shuttle buses in public transport. An image of an exemplary HAV was presented to the participants. They were also informed about the setup and features of the RO's prototypical workplace for remote assistance. Next, the experimenter invited participants to take a seat in front of the workplace.

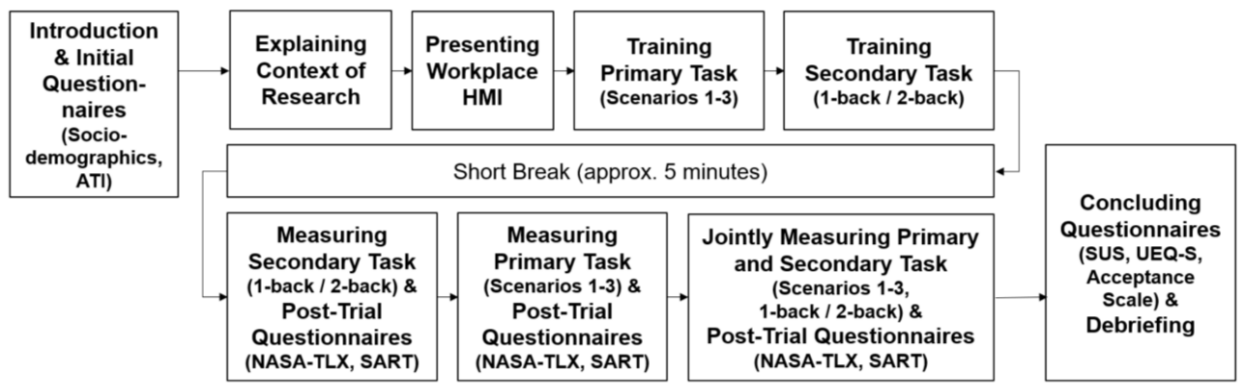

Figure 4. Overview of the study procedure.

Subsequently, participants were asked adjust their swivel chair so they could see every screen well and were able to reach all input devices, i.e., mouse, keyboard, and the tablet computer for administering questionnaires. The experimenter described the features of the workplace screen by screen. Afterwards, participants were encouraged to familiarize themselves with the workplace HMI independently by closely looking at all the screens, clicking around and learning about the implemented features. Once they were confident to have acquired a general understanding of the structure and features of the workplace, they were instructed to notify the experimenter. All participants did this within 5 to 10 minutes. Next, they were guided through the task completion process of the three scenarios (primary tasks, see also section 2.3.1) by the experimenter who commented on each step. The experimenter ensured that participants were able to understand the sequence of actions and possible interactions in all scenarios. Participants were invited to ask questions. Next, they were familiarized with the secondary task, the n-back task (Kirchner, 1958). A tablet computer located right next to the touchscreen (see Figure 2) visually and auditorily presented a digit from 1 to 9 in intervals of 5 seconds. The participants were given an example each for both variations of the task (1-back and 2-back) and underwent a trial each to familiarize themselves with the task until they felt confident with it. A short break of approx. 5 minutes concluded the training block.

In the first experimental phase that measured the baseline performance in the secondary task, participants completed one trial for both variants of the secondary task (1-back, 2-back) in a balanced order. After each of the two trials, they filled in the NASA-TLX questionnaire. In the second experimental phase that measured the baseline performance in the primary task, participants completed each of the three scenarios of the primary task (1, 2, 3) in a balanced order. Before carrying out the tasks, participants were instructed to make the journey as smooth, quick, and seamless as possible for the passengers of the HAV. Also, they were reminded of their responsibility for the passengers' safety that can only be fulfilled appropriately if close attention was directed to all the information presented on the screens and recommendations with utmost care. Participants completed the NASA-TLX and SART after each trial. Subsequently, participants completed the joint data collection of the primary and secondary task with the same n-back variant. Two blocks (1-back, 2-back) of three trials each (Scenario 1, 2, 3) were administered in counterbalanced order. Again, participants completed the NASA-TLX and SART after each of the six trials. Participants were instructed to give priority to completing the primary task but, at the same time, not to neglect the secondary task because failure to do so would disable the supervised HAV, resulting in passenger dissatisfaction. In each trial, participants performed the secondary task alone in the first 25 seconds of automated driving before the primary task was presented and had to be resolved by the participants. After completing the primary task, the secondary task lasted until 30 n-back comparisons were carried out.

Finally, participants filled in the questionnaires assessing usability, user experience, and acceptance, and were encouraged to provide remarks on the HMI and the study overall. The whole procedure took about 2.5 hours.

## 3  Results

As the study was conducted using a within-subject design, a repeated-measures analysis of variance (RM-ANOVA) was applied to determine the influence of primary and secondary task condition on the outcome variables presented above. Since the Mauchly (1940) sphericity test indicated a violation of sphericity in some cases, the Greenhouse-Geisser (1959) correction was applied in all reported RM-ANOVA results. In addition to the RM-ANOVA, post-hoc pairwise comparisons with Bonferroni (1936) correction were performed to identify significant differences between specific groups.

### 3.1  *Performance (H1)*

To test Hypothesis 1.1 to 1.3, multiple statistical procedures were used. To test whether participants required more time to react to an incoming notification under varying levels of cognitive demand (H1.1), a 3×3 RM-ANOVA was computed. The descriptive statistics regarding task reaction time are presented in Table 1. The main effect of primary task condition (scenario) on task reaction time was not significant, $F(2, 66) = 0.798$, $p = .448$, $η² = .024$ (Figure 5). There was also no main effect of the secondary task condition on task reaction time, $F(2, 66) = 3.178$, $p = .063$, $η² = .088$. Thus, induced cognitive load did not affect reaction times to incoming notifications. The respective hypothesis (H1.1) could not be accepted. There was no significant interaction effect between primary task condition and secondary task condition on task reaction time either, $F(4, 132) = .597$ $p = .612$, $η² = .018$. As shown in Table 2, no significant differences were yielded by post-hoc pairwise comparisons.

Table 1. Descriptive statistics of task reaction time in seconds by primary task (scenario) and condition of secondary task.

| Primary Task (Scenario) | Condition of Secondary Task | M | SD | CI (LL)[1] | CI (UL)[1] |
|---|---|---|---|---|---|
| 1 | None | 7.08 | 3.93 | 5.71 | 8.45 |
|   | 1-back | 6.18 | 2.37 | 5.36 | 7.01 |
|   | 2-back | 6.16 | 2.16 | 5.41 | 6.91 |
| 2 | None | 7.55 | 4.63 | 5.94 | 9.17 |
|   | 1-back | 6.22 | 1.81 | 5.59 | 6.85 |
|   | 2-back | 6.18 | 2.44 | 5.33 | 7.03 |
| 3 | None | 6.36 | 2.54 | 5.47 | 7.25 |
|   | 1-back | 6.30 | 2.71 | 5.36 | 7.25 |
|   | 2-back | 5.92 | 2.12 | 5.18 | 6.66 |

*Notes.* [1] CI = 95% confidence interval, LL = lower limit, UL = upper limit.

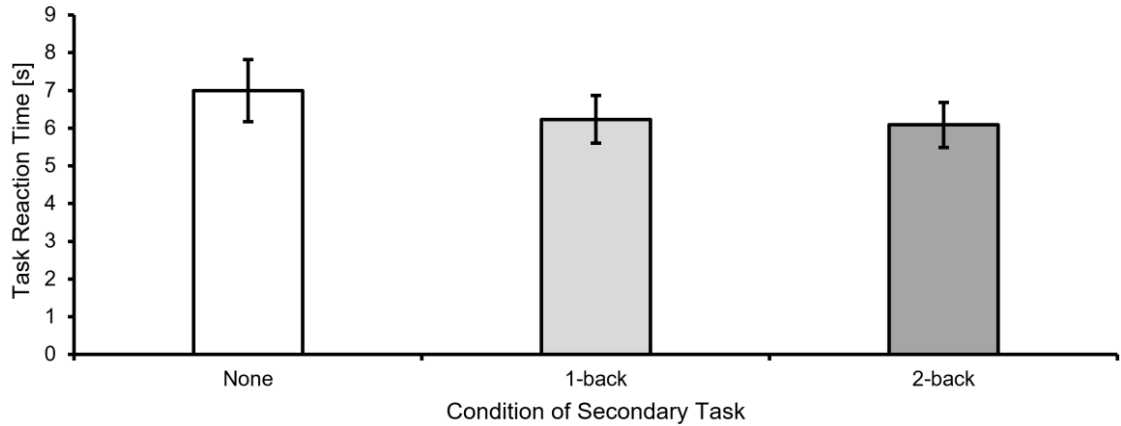

Figure 5. Means of task reaction time by condition of secondary task. Bars indicate 95% confidence intervals.

Table 2. Pair-wise comparisons of means between conditions of secondary task regarding task reaction time in seconds (Bonferroni correction applied).

| Condition of Secondary Task | $M_{diff}$[1] | $SE$[2] | $p$ | CI (LL)[3] | CI (UL)[3] |
|---|---|---|---|---|---|
| 1-back – none | -0.76 | 0.43 | .254 | -1.84 | 0.32 |
| 2-back – none | -0.91 | 0.45 | .149 | -2.04 | 0.22 |
| 2-back – 1-back | -0.15 | 0.26 | 1.000 | -0.80 | 0.51 |

*Notes.* [1] $M_{diff}$ = Difference of means between respective conditions; [2] SE = Standard error; [3] CI = 95% confidence interval, LL = lower limit, UL = upper limit.

Another 3×3 RM-ANOVA examined whether participants needed more time from the acceptance of the notification to the completion of the task at increasing levels of cognitive demand (H1.2). The descriptive statistics regarding task completion time are presented in Table 3. A significant main effect of primary task condition on task completion time was found, $F(2, 66) = 82.814$, $p < .001$, $\eta^2 = .715$. Post-hoc pairwise comparisons (Table 4) revealed that Scenario 1 took participants significantly less time to complete than Scenarios 2 and 3. This is a direct consequence of the task design, particularly due to the number of steps required to resolve it and the kind of input the RO had to provide (see section 4.1 for details). Also, a significant main effect of secondary task condition on task completion time was found, $F(2, 66) = 7.663$, $p = .002$, $\eta^2 = .188$ (Figure 6). The more cognitive load was induced by the secondary task, the longer it took participants to complete the task. Therefore, hypothesis H1.2 was accepted. There was no significant interaction effect between primary task condition and secondary task condition on task completion time, $F(4, 132) = 1.784$ $p = .152$, $\eta^2 = .051$. As shown in Figure 6, post-hoc pairwise comparisons yielded a significant difference between the 2-back and the 1-back secondary task conditions ($M_{diff} = 6.78$, $p < .001$) but not between 2-back and no secondary task. See Table 4 and Table 5 for all post-hoc comparisons.

Table 3. Descriptive statistics of task completion time in seconds by primary task (scenario) and condition of secondary task.

| Primary Task (Scenario) | Condition of Secondary Task | $M$ | $SD$ | CI (LL)[1] | CI (UL)[1] |
|---|---|---|---|---|---|
| 1 | None | 33.42 | 15.71 | 27.94 | 38.90 |
|   | 1-back | 29.43 | 8.46 | 26.48 | 32.38 |
|   | 2-back | 33.96 | 12.66 | 29.54 | 38.38 |
| 2 | None | 50.76 | 13.79 | 45.95 | 55.57 |
|   | 1-back | 47.16 | 11.38 | 43.19 | 51.13 |
|   | 2-back | 54.25 | 17.44 | 48.17 | 60.33 |
| 3 | None | 46.29 | 12.34 | 41.98 | 50.59 |
|   | 1-back | 45.32 | 9.18 | 42.12 | 48.52 |
|   | 2-back | 54.05 | 16.61 | 48.25 | 59.84 |

*Notes.* [1] CI = 95% confidence interval, LL = lower limit, UL = upper limit.

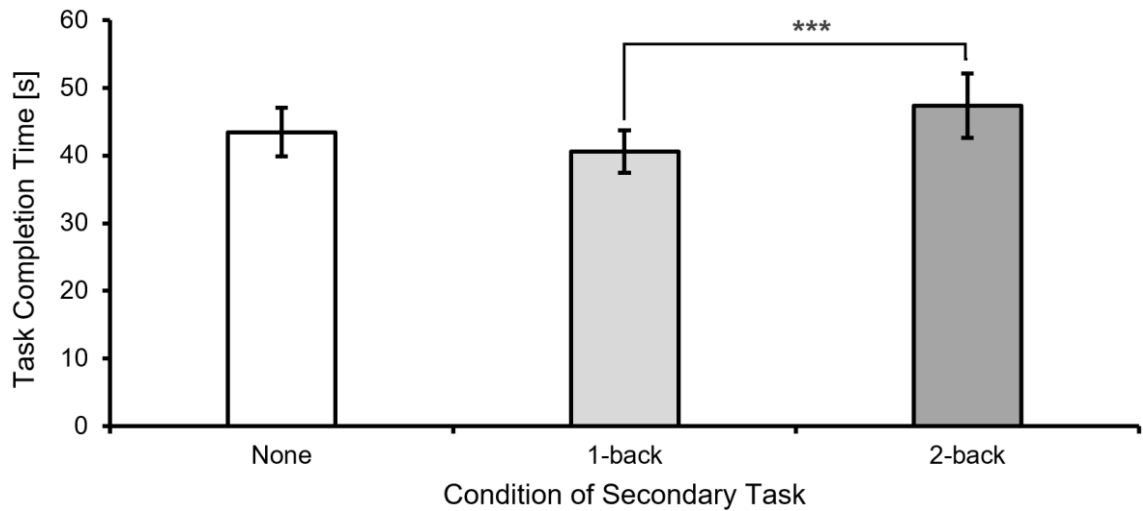

Figure 6. Means of task completion time of task by condition of secondary task. Bars indicate 95% confidence intervals. * $p < .05$, ** $p < .01$, *** $p < .001$.

Table 4. Pair-wise comparisons of means between primary tasks regarding task completion time in seconds (Bonferroni correction applied).

| Primary Task (Scenario) | $M_{diff}$[1] | $SE$[2] | $p$ | CI (LL)[3] | CI (UL)[3] |
|---|---|---|---|---|---|
| 1 – 2 | -18.45 | 1.54 | < .001 | -22.34 | -14.57 |
| 1 – 3 | -16.28 | 1.45 | < .001 | -19.93 | -12.64 |
| 2 – 3 | 2.17 | 1.71 | .635 | -2.13 | 6.47 |

*Notes.* [1] $M_{diff}$ = Difference of means between respective conditions; [2] SE = Standard error; [3] CI = 95% confidence interval, LL = lower limit, UL = upper limit.

Table 5. Pair-wise comparisons of means between conditions of secondary task regarding Task completion time in seconds (Bonferroni correction applied).

| Condition of Secondary Task | $M_{diff}$[1] | $SE$[2] | $p$ | CI (LL)[3] | CI (UL)[3] |
|---|---|---|---|---|---|
| 1-back – none | -2.85 | 1.57 | .235 | -6.81 | 1.11 |
| 2-back – none | 3.93 | 2.12 | .217 | -1.41 | 9.27 |
| 2-back – 1-back | 6.78 | 1.46 | < .001 | 3.10 | 10.46 |

*Notes.* [1] $M_{diff}$ = Difference of means between respective conditions; [2] SE = Standard error; [3] CI = 95% confidence interval, LL = lower limit, UL = upper limit.

Finally, a third 3×3 RM-ANOVA examined whether participants' numbers of correct n-back comparisons decreased at increasing levels of cognitive demand (H1.3). The descriptive statistics regarding number of correct n-back comparisons are presented in Table 6. As shown in Figure 7 and Table 7, a significant main effect of secondary task condition on number of correct n-back comparisons was found, $F(1, 33) = 63.440$, $p < .001$, $\eta^2 = .658$. That means that in the secondary condition that induced a higher cognitive load, significantly less correct n-back comparisons were made. Therefore, hypothesis H1.3 was accepted. There was no significant main effect of primary task condition on number of correct n-back comparisons, $F(2, 66) = 1.885$, $p < .160$, $\eta^2 = .054$. In addition, the interaction effect between primary task condition and secondary task condition on number of correct n-back comparisons was not significant, $F(2, 55) = 1.250$, $p = .283$, $\eta^2 = .036$.

Table 6. Descriptive statistics of numbers of correct n-Back comparisons in seconds by primary task (scenario) and condition of secondary task.

| Primary Task (Scenario) | Condition of Secondary Task | $M$ | $SD$ | CI (LL)[1] | CI (UL)[1] |
|---|---|---|---|---|---|
| 1 | 1-back | 29.82 | 0.39 | 29.69 | 29.96 |
|   | 2-back | 28.50 | 1.13 | 28.10 | 28.90 |
| 2 | 1-back | 29.68 | 0.59 | 29.47 | 29.88 |
|   | 2-back | 28.38 | 1.60 | 27.83 | 28.94 |
| 3 | 1-back | 29.71 | 0.68 | 29.47 | 29.94 |
|   | 2-back | 27.91 | 2.05 | 27.20 | 28.63 |

*Notes.* [1] CI = 95% confidence interval, LL = lower limit, UL = upper limit.

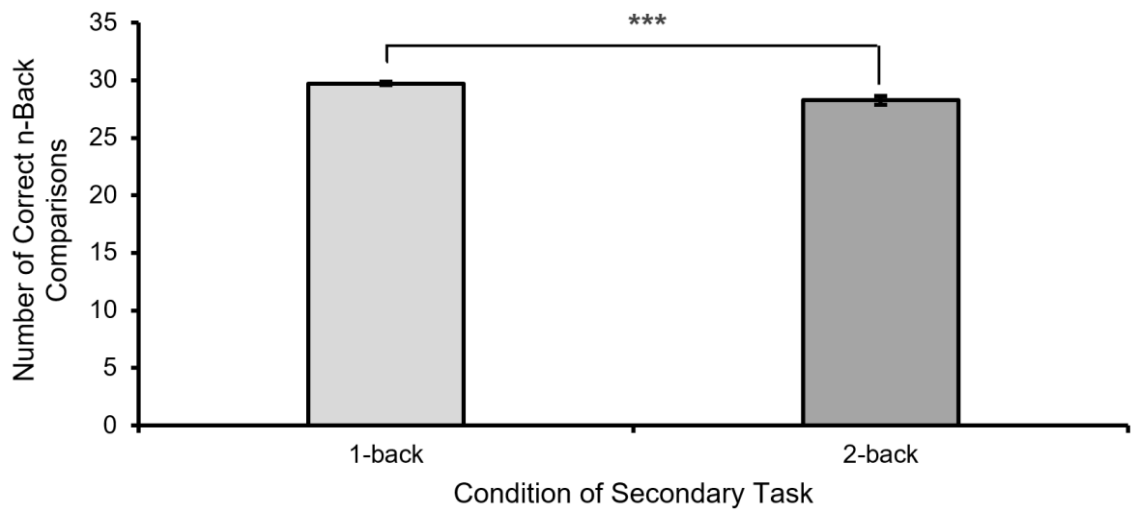

Figure 7. Means of number of correct n-back comparisons by condition of secondary task. Bars indicate 95% confidence intervals. * $p < .05$, ** $p < .01$, *** $p < .001$.

Table 7. Pair-wise comparisons of means between conditions of secondary task regarding number of correct n-back comparisons (Bonferroni correction applied).

| Condition of Secondary Task | $M_{diff}$[1] | $SE$[2] | $p$ | CI (LL)[3] | CI (UL)[3] |
|---|---|---|---|---|---|
| 2-back – 1-back | -1.47 | 0.18 | < .001 | -1.85 | -1.09 |

*Notes.* [1] $M_{diff}$ = Difference of means between respective conditions; [2] SE = Standard error; [3] CI = 95% confidence interval, LL = lower limit, UL = upper limit.

### 3.2 Situation Awareness (H2)

Hypothesis 2 examined whether ratings of situation awareness (SA) on the SART scale decrease at increasing levels of cognitive demand. Again, a 3×3 ANOVA with repeated measures was conducted. The descriptive statistics regarding SART scores are presented in Table 8. Primary task did not significantly affect participants SART score, $F(2, 66) = 0.639$, $p = .531$, $\eta^2 = .019$. However, there was a main effect of secondary task condition on participants' SART score, $F(2, 66) = 27.819$, $p < 0.001$, $\eta^2 = .457$.

Globally, a higher induced cognitive load was therefore associated with a lower SART score. Consequently, the respective hypothesis that subjective situation awareness degraded as workload increased was accepted (Figure 8). The interaction effect between primary task condition and secondary task condition on SART score was not significant, $F(4, 132) = 1.116$, $p = .349$, $\eta^2 = .033$. Pairwise post-hoc comparisons yielded significant differences between the 2-back and the 1-back secondary task conditions as well as between the 2-back and the no secondary task condition but not between 1-back and no secondary task (Table 9).

Table 8. Descriptive statistics of SART scores (Taylor, 1990) for subjective situation awareness (low: -5 – high: 13) by primary task (scenario) and condition of secondary task.

| Primary Task (Scenario) | Condition of Secondary Task | $M$ | $SD$ | CI (LL)[1] | CI (UL)[1] |
|---|---|---|---|---|---|
| 1 | None | 7.78 | 2.00 | 7.09 | 8.48 |
|   | 1-back | 7.24 | 1.83 | 6.60 | 7.87 |
|   | 2-back | 5.93 | 2.08 | 5.20 | 6.66 |
| 2 | None | 7.67 | 1.91 | 7.00 | 8.33 |
|   | 1-back | 7.04 | 2.02 | 6.34 | 7.74 |
|   | 2-back | 6.10 | 1.85 | 5.45 | 6.74 |
| 3 | None | 7.42 | 1.93 | 6.75 | 8.09 |
|   | 1-back | 7.15 | 1.82 | 6.51 | 7.78 |
|   | 2-back | 5.89 | 2.02 | 5.19 | 6.60 |

*Notes.* [1] CI = 95% confidence interval, LL = lower limit, UL = upper limit.

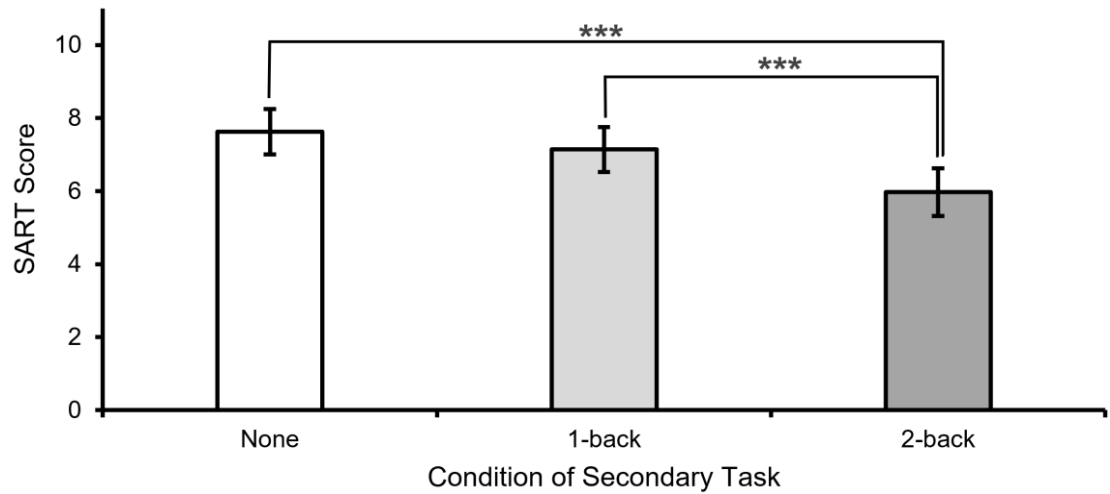

Figure 8. Means of SART scores (Taylor, 1990) for subjective situation awareness by condition of secondary task (low: -5 – high: 13). Bars indicate 95% confidence intervals.  * $p < .05$, ** $p < .01$, *** $p < .001$.

Table 9. Pair-wise comparisons of means between conditions of secondary task regarding SART scores for subjective situation awareness (Bonferroni correction applied).

| Condition of Secondary Task | $M_{\text{diff}}$[1] | $SE$[2] | $p$ | CI (LL)[3] | CI (UL)[3] |
|---|---|---|---|---|---|
| 1-back – none | -0.48 | 0.23 | 0.124 | -1.058 | 0.091 |
| 2-back – none | -1.65 | 0.25 | < .001 | -2.283 | -1.017 |
| 2-back – 1-back | -1.17 | 0.20 | < .001 | -1.674 | -0.660 |

*Notes.* [1] $M_{\text{diff}}$ = Difference of means between respective conditions; [2] SE = Standard error; [3] CI = 95% confidence interval, LL = lower limit, UL = upper limit.

### 3.3 Workload (H3)

Hypothesis 3 tested whether participants' reported ratings of workload on the NASA-TLX questionnaire increase at increasing levels of cognitive demand. The descriptive statistics regarding NASA-TLX scores are presented in Table 10. A 3×3 RM-ANOVA revealed a main effect of primary task on NASA-TLX score, $F(2, 64) = 3.748$, $p = .041$,

η² = .105. That means that among the primary tasks, subjective workload differed significantly. Post-hoc pairwise comparisons (Table 11) revealed, however, that only between Scenario 1 and Scenario 2 workload was experienced significantly differently, not between any of the other pairs of scenarios. Additionally, the main effect of secondary task condition on NASA-TLX score reached significance, $F(2, 64) = 72.767$, $p < .001$, η² = .695 (see Figure 9). Hence, the higher the cognitive load induced by the secondary task was, the higher the perceived workload was. The hypothesis that an elevated induced cognitive load leads to increased perceived workload was therefore accepted. Post-hoc comparisons were significant between all conditions (Table 12). No effect of the interaction between primary task condition and secondary task condition on NASA-TLX score was found, $F(4, 128) = 0.257$, $p = .877$, η² = .008.

Table 10. Descriptive statistics of NASA-TLX scores (Hart & Staveland, 1988) for subjective workload (low: 1 – high: 21) by primary task (scenario) and condition of secondary task.

| Primary Task (Scenario) | Condition of Secondary Task | $M$ | $SD$ | CI (LL)[1] | CI (UL)[1] |
|---|---|---|---|---|---|
| 1 | None | 6.35 | 1.98 | 5.66 | 7.04 |
|   | 1-back | 8.52 | 2.44 | 7.67 | 9.38 |
|   | 2-back | 10.11 | 2.38 | 9.28 | 10.94 |
| 2 | None | 6.72 | 2.00 | 6.02 | 7.41 |
|   | 1-back | 8.78 | 2.38 | 7.95 | 9.61 |
|   | 2-back | 10.30 | 2.42 | 9.46 | 11.14 |
| 3 | None | 6.57 | 1.80 | 5.93 | 7.21 |
|   | 1-back | 8.81 | 2.36 | 7.99 | 9.63 |
|   | 2-back | 10.55 | 2.36 | 9.73 | 11.38 |

*Notes.* [1] CI = 95% confidence interval, LL = lower limit, UL = upper limit.

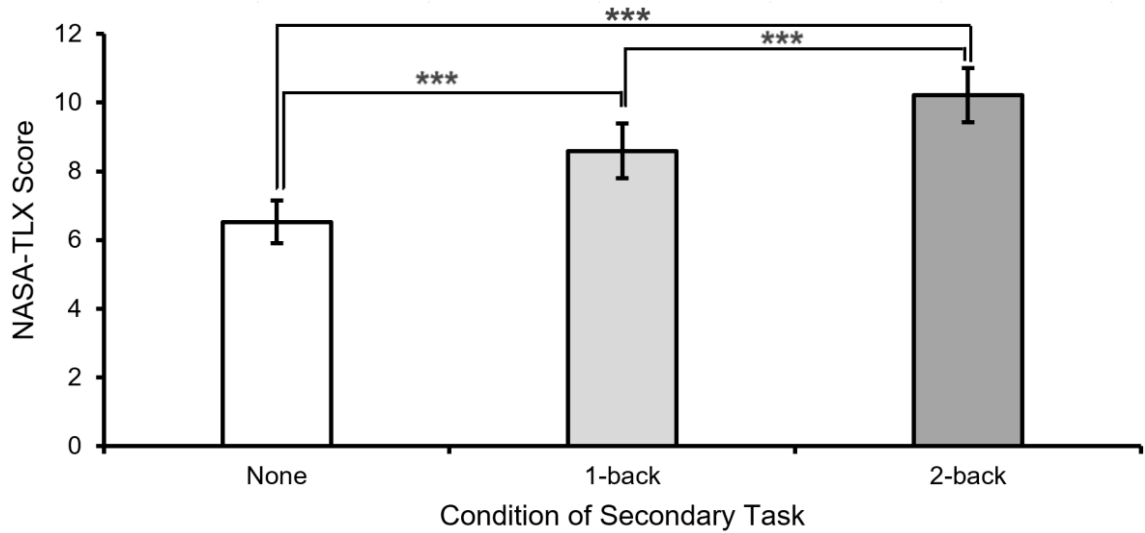

Figure 9. Means of NASA-TLX scores for subjective workload (low: 1 – high: 21) by condition of secondary task. Bars indicate 95% confidence intervals. * $p < .05$, ** $p < .01$, *** $p < .001$.

Table 11. Pair-wise comparisons of means between primary tasks (scenarios) regarding NASA-TLX scores for subjective workload (low: 1 – high: 21; Bonferroni correction applied).

| Primary Task (Scenario) | $M_{diff}$[1] | $SE$[2] | $p$ | CI (LL)[3] | CI (UL)[3] |
|---|---|---|---|---|---|
| 1 – 2 | -0.27 | 0.09 | .013 | -0.09 | 1.06 |
| 1 – 3 | -0.33 | 0.14 | .086 | 1.02 | 2.28 |
| 2 – 3 | -0.06 | 0.15 | 1.000 | 0.66 | 1.67 |

*Notes.* [1] $M_{diff}$ = Difference of means between respective conditions; [2] SE = Standard error; [3] CI = 95% confidence interval, LL = lower limit, UL = upper limit.

Table 12. Pair-wise comparisons of means between conditions of secondary task regarding NASA-TLX scores for subjective workload (low: 1 – high: 21; Bonferroni correction applied).

| Condition of Secondary Task | $M_{diff}$[1] | $SE$[2] | $p$ | CI (LL)[3] | CI (UL)[3] |
|---|---|---|---|---|---|
| 1-back – none | 2.07 | 0.33 | < .001 | -1.06 | 0.09 |
| 2-back – none | 3.69 | 0.33 | < .001 | -2.28 | -1.02 |

| Condition of Secondary Task | $M_{\text{diff}}$[1] | $SE$[2] | $p$ | CI (LL)[3] | CI (UL)[3] |
|---|---|---|---|---|---|
| 2-back – 1-back | 1.62 | 0.25 | < .001 | -1.67 | -0.66 |

*Notes.* [1] $M_{\text{diff}}$ = Difference of means between respective conditions; [2] SE = Standard error; [3] CI = 95% confidence interval, LL = lower limit, UL = upper limit.

## 3.4 Questionnaires

In order to evaluate the workplace HMI overall, questionnaires regarding user-related variables including usability, user experience and acceptance were administered after all task trials had been completed. This part of the study had exploratory character to understand if the designed HMI was user-friendly beyond the scope of specific interactions investigated during the scenarios. Consequently, no hypotheses were stated a priori.

First, the System Usability Scale (SUS; Brooke, 1996) yielded very good usability ratings ($M = 76.25$, $SD = 11.87$, on a scale from 0: very poor usability to 100: flawless usability). This score rates between the adjective ratings "good" ($M = 72.8$) and "excellent" ($M = 85.6$) as resulting from Bangor et al.'s (2008) empirical validation of the verbal interpretation of SUS scores. The interaction design between users and the investigated HMI was therefore regarded positive.

Second, user experience was measured with the User Experience Questionnaire Short Version (UEQ-S). The questionnaire consists of two subscales, the pragmatic and the hedonic scale. While the former scale pertains to a concept similar to usability, the latter focuses on the emotional component of user experience. Both subscales and the complete scale were tested against the arithmetic center of the scale, 0. This approach was based on the assumption that the center of a scale represents a conceptual average value, e.g., a medium extent of usability. Table 13 shows the descriptive and test results. User experience is rated significantly higher than the arithmetic scale mean on both the complete scale and the pragmatic scale. This finding indicates that participants were satisfied with their interactions with the workplace HMI. Results on the hedonic scale did not differ significantly from the mean, suggesting average emotional experiences with the HMI.

Table 13. Subjective user experience of the prototypical remote operation workplace measured with UEQ-S.

| Scale | $M_{emp}$ | $SD_{emp}$ | CI (LL)[1] | CI (UL)[1] | $M_{crit}$[2] | t | p |
|---|---|---|---|---|---|---|---|
| Overall | 0.38 | 0.50 | 0.21 | 0.56 | 0 | 4.502 | < .001 |
| Pragmatic | 0.63 | 0.46 | 0.47 | 0.78 | 0 | 7.966 | < .001 |
| Hedonic | 0.14 | 0.95 | -0.19 | 0.47 | 0 | 0.861 | .395 |

*Notes.* User Experience Questionnaire – Short (UEQ-S; Schrepp, Hinderks & Thomaschewski, 2017); low: 1 – high: 7 (center : 4). [1] CI = 95% confidence interval, LL = lower limit, UL = upper limit. [2] $M_{crit}$ = critical mean to test empirical mean ($M_{emp}$).

Third, acceptance of the HMI was measured with Van der Laan et al.'s (1997) Acceptance Scale. Similar to the UEQ-S, it consists of two subscales, one of which focuses on the usability, or usefulness, to use the term that the authors of the questionnaire used, while the other one centers around the emotional quality of the HMI. As shown in Table 14, it was found that both the overall scale mean and the means of the subscales usefulness and satisfaction were significantly higher ($p < .001$) than the center of the scale, 3, indicating that acceptance, usefulness and satisfaction with the prototype were considered to be above average. These findings are very similar to the UEQ-S results, with the difference that satisfaction was rated more favorably on the satisfaction subscale of the Acceptance Scale than the pragmatic quality in the UEQ-S.

Table 14. Subjective acceptance of the prototypical remote operation workplace measured with the Acceptance Scale.

| Scale | $M_{emp}$ | $SD_{emp}$ | CI (LL)[1] | CI (UL)[1] | $M_{crit}$[2] | $t$ | $p$ |
|---|---|---|---|---|---|---|---|
| Overall | 4.02 | 0.58 | 3.82 | 4.22 | 3 | 10.181 | < .001 |
| Usefulness | 3.98 | 0.60 | 3.77 | 4.19 | 3 | 9.565 | < .001 |
| Satisfaction | 4.06 | 0.58 | 3.86 | 4.26 | 3 | 10.639 | < .001 |

*Notes.* Van der Laan Acceptance Scale (Van der Laan, Heino & De Waard, 1997); low: 1 – high: 5 (center: 3); [1] CI = 95% confidence interval, LL = lower limit, UL = upper limit. [2] $M_{crit}$ = critical mean to test empirical mean ($M_{emp}$).

# 4 Discussion

The goal of this study was to evaluate a prototypical workplace for the remote assistance of highly automated vehicles (HAVs) regarding performance, situation awareness (SA), workload, and other user-related outcomes. Serving as representatives of the future user group of remote operators, participants resolved scenarios that were considered relevant routine tasks in HAV remote operation as listed by Kettwich et al. (2022) using the prototypical workplace's HMI. Furthermore, a secondary task was added to elevate the participants' workload, thus simulating the execution of parallel tasks or distractions. Three hypotheses were postulated regarding performance, SA and workload of participants while resolving three scenarios with the workplace HMI.

## 4.1 Results on H1 (Performance)

The first hypothesis assumed that participants showed lower performance at increasing levels of cognitive demand. This hypothesis was partially accepted: a significant main effect of secondary task condition, which was used as a proxy to systematically vary cognitive demand, was reached for task completion time but not for task reaction time. This finding suggests that induced cognitive load has a negative effect on *processing* a task in a timely manner (H1.2) but not on the response time that a participant needs to *react* to an incoming notification (H1.1). This finding can be explained with Wickens' (2002) multiple resource theory: The process of resolving the relatively simple primary task poses a considerable cognitive demand onto the operator, hence consuming a share of the pool of cognitive resources that is available to them. Cognitive demand induced by the secondary task draws from the same pool and therefore competes with the primary task's execution for cognitive resources. This diminishes the supply of cognitive resources for completing the primary task, resulting in a longer task completion time. Since this effect occurs already in simple and well-trained routine tasks that participants were subjected to in this study, it is probable that the RO's workload will increase as tasks and interactions become more complex and novel. It is therefore questionable whether additional tasks beyond remotely supporting HAVs can be assigned to ROs.

In contrast to completing a task, accepting a task is assumed to be cognitively less demanding. The additional demand posed by the secondary task does not deplete the pool of resources as strongly as processing it as it is a simple procedure that does not require abundant cognitive resources. Moreover, the pattern to accept the request is identical across tasks. Thus, instead of high-level cognitive mechanisms like working memory, more basal reaction times to visual stimuli might influence the results regarding task reaction time. They do not depend on a common pool of cognitive resources but are separate cognitive processed that might be explained as phenomena of attention distribution, e.g., by Wickens et al.'s (2003) SEEV model. According to this model, both bottom-up and top-down mechanisms of cognitive processing come into play when directing attention to stimuli. Specifically, the SEEV model claims that a stimulus's likelihood to attract attention is influenced by its *salience*, the *effort* it takes to perceive it, the *expectancy* to perceive it, and the *value* the perceiver assigns to it. Furthermore, reaction time to a stimulus may be influenced by factors such as intensive training (Barrett et al., 2020). Consequentially, reacting to incoming notifications is not compromised by inducing more cognitive load, resulting in similar reaction times between the secondary task conditions.

Concerning the secondary task, it was hypothesized that the number of correct n-back comparisons decreases at increasing levels of cognitive demand (H1.3). This hypothesis was accepted, demonstrating the successful induction of cognitive load that diminished the participants' performances at the secondary task. The finding indicated that increasing the level of *n* does indeed result in a higher reported workload. The n-back task can therefore be considered valid at least as an approximation for additional work-related or unrelated tasks that occur in a real-world setting. An example for an additional task is being responsible for services other than core remote operation, such as communicating with passengers on board of the assisted HAV. This finding is in favor of splitting up core remote assistance tasks, such as ensuring the HAV's onward travel, from surrounding tasks, such as passenger communication, into different roles. Separate roles could help avoid cognitive overload in ROs in situations where several tasks would have to be executed simultaneously, ensuring safe and efficient operations.

The main effect of primary task on task completion time results from the finding that Scenario 1 took participants significantly less time to complete than Scenarios 2 and 3. This effect can be traced back to the differing lengths of the displayed scenarios and the kind of required interaction between participant and workplace HMI. For instance, in Scenario 1 only clearance had to be given whereas in Scenarios 2 and 3 longer, multi-step interactions were required to resolve the task.

## 4.2 Results on H2 (Situation Awareness)

It was postulated in the second hypothesis that participants' ratings of situation awareness (SA) decrease at increasing levels of cognitive demand. This was also found in the collected data: The level of reported SA slightly deteriorated as cognitive load increased. The finding means that a higher cognitive load degrades subjective SA, even in routine and well-trained tasks in which participants could be expected to maintain SA due to learning effects. The finding has implications on the design of the interaction between RO and HAV: Keeping the RO's workload at a manageable level may help them maintain a sufficient degree of SA. To ensure this, one way could be for the RO to focus their cognitive capabilities on resolving the HAV's request solely, without attending to other tasks, at least while processing the request. The result that executing other tasks in parallel to a main task lowers SA is consistent with existing literature. Merat et al. (2010) found in a driving simulator study that SA, measured by participants' responses to critical incidents, was negatively affected by an auditory secondary task. In a similar vein, drivers who had to navigate the menu of a driver information system as a visual secondary task while driving in a simulator had a significantly lower SA (Wulf et al., 2013). Comparable findings that show an association between workload and SA were also made in other domains: In the aviation domain, Endsley and Rodgers (1998) found a significant relationship between workload and several indicators of SA.

Regardless of the condition of primary or secondary task, SA scores ranged in a medium to high area. They were significantly above the arithmetic center of the scale in all levels. Therefore, it can be assumed that the HMI design is robust against SA degradation even when additional workload is induced, regardless of the scenario – at least for scenarios similar to the ones investigated in this study, i.e., routine and rather well-trained scenarios. However, whether this will hold true in more complex, less trained scenarios is for future research to examine. Finally, it is noteworthy that no significant difference was found between the 1-back condition and the no secondary task condition for the SART score. This result gives rise to the notion that a floor effect showed in the 1-back condition, with this secondary task condition not impeding SA more strongly than the condition without any secondary task.

## 4.3  *Results on H3 (Workload)*

The third hypothesis assumed that reported ratings of workload increase at rising levels of cognitive demand. The conducted RM-ANOVA confirmed this hypothesis. The finding indicates that the cognitive demand induced by a secondary n-back task fulfilled its purpose, actually increasing the perceived workload in participants. Thus, the task proved effective to reach the intended goal of artificially elevating workload to emulate working on multiple tasks simultaneously. Workload means did not transgress the center of the scale in any condition, ranging between 6 and 11 on a scale from 1 to 21. This finding suggests a low-to-medium perceived workload in all conditions of the secondary task. The simplicity and routine character of the primary task may have contributed to a low workload baseline. Nevertheless, effects of the additional cognitive load induced via a secondary task on perceived workload showed statistical significance.

The identified effect of cognitive secondary task load is in line with research in automotive HMI research that used the n-back task as a secondary task to induce cognitive load. Liang and Pitts (2019), for instance, reported that workload measured by the NASA-TLX questionnaire was significantly elevated when the difficulty level of the n-back task was increased. The study measured participants' performance in a simulated driving task supported by a lane keeping system.

In addition to the significant main effect of secondary task, a significant main effect of primary task on reported workload was revealed. This finding is a logical consequence of the varying degree of complexity and demand inherent in the different scenarios. For instance, when participants only had to give clearance to a maneuver that the driving automation suggested as in Scenario 1 workload was low, while Scenarios 2 and 3 required more complex interactions, such as drawing waypoints and selecting a new route on a map and resulted in higher workload. Designing an HMI for the RO therefore needs to take into account how the respective tasks are structured and how much complexity and workload they entail in order to determine whether and which additional tasks may be assigned to the RO. Since even simple tasks with a moderate level of induced additional cognitive load led to increased perceived workload in this study, it needs to be carefully and critically analyzed whether more workload is tolerable. It is advisable to follow a conservative approach, entrusting the RO with a small task set initially and adding new tasks incrementally and cautiously under constant observation of their impact on safety and performance until an ideally balanced task load is established.

## *4.4  Results on Questionnaires*

In addition to the variables that were directly linked to hypotheses, the three user-related outcome measures usability, user experience and acceptance were collected to evaluate the workplace HMI overall. Intentionally, no hypotheses were specified beforehand as the goal was to capture the users' impressions with and experiences of the HMI in an exploratory fashion.

First, usability was found to be in a good to excellent range as measured with SUS. In a similar vein, both the usefulness dimension of the acceptance scale and the pragmatic scale of UEQ-S showed results significantly above the arithmetic centers of the scales. Thus, all the indicators suggest a good level of usability. This is in line with the objectives of the designed prototype: ensuring basic functionalities and interactions between the workplace and the RO.

Second, user experience scales were administered to capture the emotional aspect of user interaction. Average assessments were given by participants on the hedonic subscale of UEQ-S but positive assessments on the satisfaction dimension of the Acceptance Scale. These more moderate-to-positive ratings relative to usability suggest that the emotional quality of the interaction is on a good way in the evaluated HMI but may be further refined. Consequently, after sufficient usability of the remote operation workplace has been demonstrated, future iterations of its HMI design should focus on improving user experience even more beyond the phase of the user's immediate interaction with it. However, the somewhat mixed results regarding user experience are comprehensible in the context of the development of the prototype, which did not have the priority to develop a system with outstanding user experience. The authors prioritized generating a prototype as a "proof of concept" to demonstrate its capability to process basic scenarios and tasks. Creating particularly positive experiences while interacting with the workplace is a quest for further refinement.

Third, the acceptance rating, operationalized as the overall Acceptance Scale, was above average, suggesting that participants could imagine to work with the prototypical workplace in general. This is an important finding as it demonstrates the willingness of participants to use the workplace HMI, a fundamental prerequisite for its deployment.

*4.5   Limitations*

Even though the approach taken in this study pursued ecological validity, it comes with four limitations.

First, the scenarios used in the study aimed at representing typical routine scenarios and the tasks related to resolving them from a RO's perspective. They are limited in number (only three scenarios were used throughout the study) and so is the range of tasks included. Inevitably, participants habituated to the tasks, rendering potential effects of novelty on performance improbable. However, the objective of the reported study was to evaluate the basic features and interaction patterns between RO and workplace. This was achieved by training and performing a set of standard tasks and scenarios deemed to be executed on a daily basis. Since acquiring new skills is based on general learning mechanisms, honing the skill to effectively assist HAVs with the evaluated workplace HMI is assumed to be gradual, hierarchical and based on gaining experience: Only after basic core tasks can be executed successfully, like those examined in this study, more complex and novel tasks may be feasible for trained RO. The authors thought it to be beneficial to pursue the same approach in user testing by initiating user studies with a limited set of routine scenarios and tasks and gradually extend this set.

Second, only a specific variant of HAV remote operation was investigated since the workplace HMI is tailored to it: remote assistance. Hence, the results reported in this study may not be directly transferable to other variants of remote operation, specifically, remote driving. It can be assumed, however, that the Human Factors investigated here, workload and SA, will be equally or even more critical in light of as-if real-time situational representations and immediate interventions, as required for remote driving. If remote driving becomes technically feasible and legally permissible, assessing it from a Human Factors perspective and proposing HMIs for it will become more relevant. In a similar vein, combining different roles with divergent tasks within a remote operation center, as for example suggested by Schrank and Kettwich (2021), will require modifications of the evaluated workplace HMIs as tasks are likely to change and diversify.

Third, the current legal situation in Germany demands that the Technical Supervisor's interaction with the supervised HAV be not time-critical. It could therefore be argued that the outcome variables task reaction time and task completion time that both measure durations of RO-HMI interactions are not an adequate measure to examine the workplace HMI's performance. However, even though this variable does not reflect the current legal status, the variables were deemed key performance indicators as efficiency is vital for systems to be economically viable. Only if handling an incident with an HAV via remote assistance is favorable time-wise, a business case that involves this technology may emerge.

Fourth, it can be argued that the secondary task used in this study, the n-back task, is somewhat artificial as it does not occur in the natural environment of an RO. Additionally, it does not directly interfere with the primary task as different sensory modalities are used, implying less interference (Wickens, 2002): While the visual modality is used for processing the primary task at the workplace HMI, the auditory modality is used for the secondary task. However, the n-back task is a reliable and commonly used method that ensures internal validity and enables systematic variation of induced cognitive load, enabling drawing steadfast conclusions about the effects of modulating cognitive load. Regarding different modalities, presenting the primary and secondary tasks on dissimilar modalities may actually be an advantage for determining the lower threshold of workload that is to be expected while conducting remote assistance tasks: Since differences in workload were measured even in different modalities at rather low levels of induced cognitive load and simple routine tasks, it is to be expected, according to Wickens's theory, that cognitive load that is induced via a secondary task on the same modality may impede the performance at the primary task even more, particularly when complex and novel tasks come into play. The outcome observed here may therefore be considered a lower boundary of workload that can be further increased. What the upper boundary of workload may be is subject to future research. Following from these findings, imposing even more responsibilities onto a RO is likely not to be beneficial for their performance and ought to be considered with caution.

Nevertheless, in order to increase ecological validity on top or in lieu of internal validity, future research could use more natural secondary tasks. An example for such a task is supporting passengers on board of the assisted vehicle over intercom, including auditory interfaces, to provoke same-modality interferences. This is argued to deplete ROs' cognitive resources even more, making it less likely they are still capable to resolve their tasks using the proposed workplace HMI.

## 4.6   Conclusions and Future Research

This study has shown that the presented user-centered HMI for a remote assistance workplace helps execute routine tasks by potential users – even when additional moderate cognitive load was induced via a secondary task. Thus, the remote assistance workplace HMI appears to be a feasible way to design the interaction between RO and HAV for supporting HAVs in routine scenarios, utilizing human cognitive skills. The proposed workplace HMI for remote operation proved capable of enabling a remote operator to provide support. This claim was supported by four central outcomes, relating to performance, SA, workload, and global evaluation measures.

First, even though induced cognitive load did have a significant impact on one of the performance indicators (task completion time but not task reaction time), perceived workload did not surpass a medium level. This finding indicated that at least for simple routine scenarios and related tasks, workload was maintained at a manageable level. It is for future research to examine whether the same holds true for more complex scenarios and interactions, particularly when they have not been encountered before.

Second, the degree of induced cognitive load had a negative main effect on the participants' perceived SA when processing routine tasks using the presented workplace HMI. That means that in this context, SA degrades as cognitive load increases. It must be noted, however, that the absolute differences among SA scores between conditions were moderate, with all scores ranging in a medium to upper level. Nevertheless, this finding bears significance as it gives rise to the expectation that in complex, hardly trained or entirely novel scenarios these effects may be more pronounced. Thus, higher cognitive load resulting from more challenging tasks is likely to impede SA even more. Indirectly, a diminished level of SA may affect performance negatively as the RO may make wrong assumptions about the current status of the traffic environment, which in turn might entail drawing wrong conclusions on how the situation will unfold. Future research may therefore shed light on the generation and maintenance of SA in more complex, novel scenarios. A systematic variation of scenario complexity may help elucidate how SA develops across varying levels of complexity.

Third, the main effect of the secondary task condition on perceived workload can be considered as passing the "manipulation check": inducing cognitive load indeed resulted in increasing perceived workload. Thus, the n-back task in the conditions 1-back and 2-back proved to have an impact on how much workload participants experienced. However, similar to SART scores, the differences between conditions were generally low. In all conditions, workload means did not exceed the arithmetic center of the scale. The variance was therefore narrow. Hence, a takeaway for future research may be increasing the spectrum of the secondary task's difficulty levels to inquire the effects of high induced cognitive load on primary task performance.

Fourth, the global evaluation of the workplace HMI produced favorable results. This is particularly true regarding the variable that pertains to the direct interaction with the HMI, usability. On three questionnaires (SUS, UEQ-S, Acceptance Scale), this variable or related concepts consistently yielded above-average ratings. The hedonic-emotional quality of the interaction was assessed slightly more modest but still sufficient in an average-to-positive range. This finding aligns with the authors' objective to give priority to a proof of concept that works for basic interactions before focusing on user-experience-related aspects. Finally, the HMI's acceptance ratings were positive. This result is meaningful because the sample was tech-savvy, receiving high scores on the Affinity for Technology Scale. The HMI lived up to the potentially higher expectations that may be posed by technologically experienced and invested participants. In addition, the sample fulfilled the educational requirements of the legally specified role of the Technical Supervisor. It is of utmost importance for the utilization of an HMI that the future group of users adopts it. The results in the Acceptance Scale provide at least hints for this claim. However, future iterations of the workplace HMI need to examine critically whether groups of people beyond the Technical Supervisor may also be capable of remotely assisting HAVs. Including a wider group of participants may thus be a goal for future research.

To conclude, when designing a workplace for the remote operator, bearing in mind the Human Factors involved and their interaction with the remote operation technology are essential to ensure safety and feasibility of the system overall. Only by applying user-centered methods, using workplaces for remote operation can become a successful approach to booster automated driving technologies and thus lay the groundwork for a more sustainable mobility.

Acknowledgments

This research received funding as part of the project GAIA-X 4 ROMS – Support and Remote Operation of Automated and Networked Mobility Services (FKZ: 19S21005C). The joint project is funded by the German Federal Ministry of Economics and Climate Protection (BMWK). Solely the authors are responsible for the content of this article.

The authors would like to express their gratitude to Hoai Phuong Nguyen, Florian Rudolph and Nils Wendorff for their steadfast support in providing the technical framework for the study, Thorben Brandt for his valuable comments to the manuscript, as well as student assistants Sarah Helweg and Fabian Reese for their assistance with the data collection.

The authors confirm that there are no relevant financial or non-financial competing interests to report.

## 5 Appendix

Table A1. Steps of interaction between remote operator (RO) and workplace HMI by scenario.

| Scenario | Steps |
|---|---|
| Scenario 1: Detected Situation Unclear | (1) A message appears on the disturbance ticker under "Incoming messages": "Detection situation unclear" including more detailed information on this disturbance.<br>(2) The RO clicks on "Accept" and thus assigns himself the responsibility for this incident.<br>(3) The message disappears under "Incoming messages" and is now displayed under "Tasks" with the status "active"<br>(4) The RO clicks on "Details" under "Tasks" and receives more detailed information about the task in a pop-up window. At the same time, the camera screens display the current video streams, the details screen displays the details of the corresponding AV, and the map screen displays an environment map with the AV centered.<br>(5) In the dialogue menu, the RO is asked the following question: "Can vehicle proceed?".<br>(6) The RO clicks on the answer options "Yes, allow further travel".<br>(7) A pop-up window appears for checking the prerequisites.<br>(8) The status of the task changes to "completed".<br>(9) An according entry is made in the operating log. |
| Scenario 2: Blocked Lane | (1) The following entry appears on the fault ticker under "Incoming messages": "Lane blocked" including more detailed information on this.<br>(2) The RO clicks on "Accept" and thus assigns himself the responsibility for this incident.<br>(3) The message disappears under "Incoming messages" and is now displayed under "Tasks" with the status "active".<br>(4) The RO clicks on "Details" under "Tasks" and receives more detailed information about the task in a pop-up window. At the same time, the camera screens display the current video streams, the details screen displays the details of the corresponding AV, and the map screen displays a map of the environment with the AV centered on it.<br>(5) 5. in a dialog menu, the RO is prompted: "Draw waypoints for bypass". At the same time, a detailed map of the AV's immediate surroundings appears on the touchscreen. Using the "Draw waypoints" tool, the RO now sets waypoints that the automation converts into a drivable alternative trajectory.<br>(6) In the dialogue menu, the RO is asked the following question: "Can the alternative pathway be used?<br>(7) The RO clicks on the answer options "Give clearance now".<br>(8) The status of the task changes to "completed". |

| Scenario | Steps |
|---|---|
| | (9) An according entry is made in the operating log. |
| Scenario 3: Rerouting | (1) The following entry appears under "Incoming messages": "Road closed" including more detailed information on this.<br>(2) The RO clicks on "Accept" and assigns themselves the responsibility for this incident.<br>(3) The message disappears under "Incoming messages" and is now displayed under "Tasks" with the status "active".<br>(4) The RO clicks on "Details" under "Tasks" and receives more detailed information about the task in a pop-up window. At the same time, the camera screens show the current video streams, the details screen shows the details of the corresponding AV, and the map screen shows an environment map with the AV centered.<br>(5) A dialogue menu prompts the RO: "Select an alternative route to bypass from the suggested routes on the touchscreen".<br>(6) At the same time, a detailed map of the AV's surroundings appears on the touchscreen showing alternative routes avoiding the closed road. In a dialogue menu on the touchscreen, the RO is prompted: "Select an alternative route to bypass from the suggested routes". In the details screen, the "Location" tab is automatically invoked. In the table of next stops, cancelled stops are indicated with "CANCELLED: [name stop]" in red font. New stops are displayed with "NEW: [name stop]" in green font.<br>(7) In the dialogue menu, the RO is asked the following question on the touch screen: "Select an alternative route to bypass from the suggested routes."<br>(8) The RO clicks on the response options "Select route now".<br>(9) At the same time, the location is displayed in the Detail Screen and the stops to be served and the stops to be cancelled are shown.<br>(10) The status of the task changes to "completed".<br>(11) An according entry is made in the operating log. |